# Electronic Transport Studies of Ag-doped Bi$_2$Se$_3$ Topological Insulator

*Shailja Sharma[1], Shiv Kumar[2], Amit Kumar[3], Kenya Shimada[2], C.S. Yadav[1*]*

[1]School of Basic Sciences, Indian Institute of Technology Mandi, Mandi-175075 (H.P.) India
[2]Hiroshima Synchrotron Radiation Center, Hiroshima University, 2-313 Kagamiyama, Higashi-Hiroshima 739-0046, Japan
[3]Graduate School of Science, Hiroshima University, Kagamiyama 1-3-1, Higashi-Hiroshima 739-8526, Japan
*Email: shekhar@iitmandi.ac.in

The structural, magnetotransport and angle-resolved photoemission spectroscopy (ARPES) of Ag-doped Bi$_2$Se$_3$ single crystals are presented. Temperature dependent resistivity exhibit metallic behaviour with slope change above 200 K for Ag-doped Bi$_2$Se$_3$. The magnetoresistance show positive quadratic dependence at low fields satisfying Kohler's rule. Hall resistivity measurement shows that electrons are dominant charge carriers. Further, these results agree well with the ARPES spectra observed at $T$ = 20 K, where the Fermi level lies inside the bulk conduction band. The Dirac point of the topological surface states (TSSs) is shifted towards higher binding energy (~ 120 meV) for Ag-doped samples as compared to pristine Bi$_2$Se$_3$.

## 1. Introduction

Topological Insulators (TIs) have been widely investigated owing to their exotic electronic properties and potential futuristic applications. TIs are a novel and exciting state of quantum matter with bulk insulating and conducting surface states, originating from band inversion due to strong spin-orbit coupling.[1, 2] Bi$_2$Se$_3$ is one of the most studied TIs with a single Dirac cone at the $\bar{\Gamma}$ point and a large direct band gap of ~ 300 meV. The existence of spin-momentum coupled topological surface states (TSS) has been experimentally confirmed through surface sensitive techniques such as angle-resolved photoemission spectroscopy (ARPES) and scanning tunneling microscopy/spectroscopy (STM/STS).[3-5] Transport properties confirming the 2D TSS in Bi$_2$Se$_3$ has been masked by the bulk dominating conductivity. This is because the Fermi level lies in the bulk conduction band, therefore, Bi$_2$Se$_3$ is intrinsically *n*-type, resulting in metallic resistivity behaviour. This *n*-type behaviour has been attributed to the deficiency of Se atoms that are difficult to control during the synthesis process.[6]

After the discovery of superconductivity in Cu$_x$Bi$_2$Se$_3$[7], chemical doping and pressure-induced superconductivity became the topic of interest for the emergence of topological superconductivity. Transition metal (Cu, Sr, Nb) doped Bi$_2$Se$_3$ show superconductivity under ambient pressure.[8-10] Furthermore, evolution of crystal structure under pressure in pristine Bi$_2$Se$_3$ results in phase transition from ambient-pressure rhombohedral to monoclinic at around 10 GPa and to cubic phase at higher pressures, where the emergence of superconductivity has been observed.[11-13] Similar structural transition study in Ag$_x$Bi$_{2-x}$Se$_3$ has been reported by Tong *et al*.[14], from rhombohedral-monoclinic-tetragonal and subsequently, emergence of superconductivity at a pressure of 11 GPa. The pressure-induced superconductivity in Ag$_x$Bi$_{2-x}$Se$_3$, where Ag atoms was substituted for Bi atoms is intended due to hole-doped Bi$_2$Se$_3$ whereas the superconductivity discussed in Cu, Sr, and Nb-doped Bi$_2$Se$_3$ is due to electron doping.[14] Subsequently, to understand the electronic structure, Yamaoka *et al*. carried out the combined study using X-ray absorption spectroscopy and density functional theory on Bi$_2$Se$_3$ and Ag$_x$Bi$_2$Se$_3$.[15] Another study by Uesugi *et al*. shows Fermi-level tuning in Ag-doped Bi$_2$Se$_3$ field-effect transistor (FET) structured devices, which reveal a change from metallic behaviour in non-doped Bi$_2$Se$_3$ to insulating nature.[16] Also, superconductivity ($T_c$ ~ 4.5 K) in Ag$_x$Bi$_2$Se$_3$ at ambient pressure has been proposed using a dynamical-mean-field theory with local density approximation.[17] More recently, local and erasable superconductivity has been observed across the interface of Ag-point contact on Bi$_2$Se$_3$ evidenced by Andreev reflection peak on the same region and drop in resistance, implying that superconductivity is possibly due to the Ag-dopants in the van der Waals gap in Bi$_2$Se$_3$.[18] Therefore, considering the above mentioned pressure-dependent and theoretical investigations, we have carried out detailed magnetotransport and ARPES studies of Ag-doped Bi$_2$Se$_3$.

The synthesis process plays an important role in deciding the possibility of Ag atoms occupying the substitution or intercalation sites. If Ag atoms substitute Bi atom as in case of Ca doping in Bi$_2$Se$_3$[6], then Ag doped should be *p*-type. However, both Ag$_x$Bi$_{2-x}$Se$_3$ and Ag$_x$Bi$_2$Se$_3$ were found to be *n*-type. In this paper, we report the electronic transport study of Ag-doped Bi$_2$Se$_3$ single crystals, synthesized using two different processes that favour (i) substitution of Ag for Bi atom as Ag$_x$Bi$_{2-x}$Se$_3$ and (ii) intercalation of Ag atoms in the interlayer spacing of Bi$_2$Se$_3$ as Ag$_x$Bi$_2$Se$_3$. A considerable change in low-temperature resistivity behaviour for Ag$_x$Bi$_{2-x}$Se$_3$ and Ag$_x$Bi$_2$Se$_3$ have been observed. Additionally, temperature dependence of resistivity behaviour for Ag-doped compounds at high temperatures (> 200 K) is quite linear compared to undoped Bi$_2$Se$_3$.

## 2. Results and discussion

### 2.1 Structural aspects

Figure 1 (a) depicts X-ray diffraction patterns on the crystals with strong reflections oriented along the (00*l*) axis confirms the single crystalline nature of the materials. Figure 1 (b) show the Raman spectra of materials at room temperature exhibiting Raman active phonon modes at ~ 33 cm$^{-1}$ ($E_g^1$), 71 cm$^{-1}$ ($A_{1g}^1$), 132 cm$^{-1}$ ($E_g^2$), and 173 cm$^{-1}$ ($A_{1g}^2$) where $E_g$ and $A_{1g}$ denote in-plane and out-of-plane atomic vibrations, respectively. The presence of four Raman modes are intrinsic to that observed in parent compound Bi$_2$Se$_3$.[19, 20] Structural parameters analysed using Rietveld refinement of powder X-ray diffraction patterns using *FullProf Suite* are presented in Fig.1 (c, d, e). The compounds crystallise in rhombohedral phase with space group $R\bar{3}m$ (#166) Bi$_2$Se$_3$. Ag$_x$Bi$_2$Se$_3$ ($x$ = 0.05, 0.10) show the phase purity and refined well with the space group #166. The variation in volume of unit cell for different Ag content compounds are presented in Table 1. The lattice constants *a* and *c* are close to that for Bi$_2$Se$_3$ (4.1375(6) Å and 28.6307(9) Å), respectively.[20] The lattice constant *c* does not are show any appreciable expansion in the case of Ag$_x$Bi$_2$Se$_3$. Ag$_{0.05}$Bi$_{1.95}$Se$_3$ shows a secondary phase formation (~ 7%) due to AgBiSe$_2$ that crystallises in hexagonal phase ($Pm\bar{3}1$) (#164). The formation of secondary phase depends upon the synthesis process where we initially took Ag, Bi and Se together. Considering the electronegativity of elements Ag (1.93), Bi (2.02), Se (2.55), it is quite possible to form AgBiSe$_2$ phase in addition to Bi$_2$Se$_3$. Since the atomic size of Ag (1.44 Å) is smaller than that of Bi (1.70 Å), the lattice mismatch is higher in Ag$_x$Bi$_{2-x}$Se$_3$ than that of Ag$_x$Bi$_2$Se$_3$.[21]

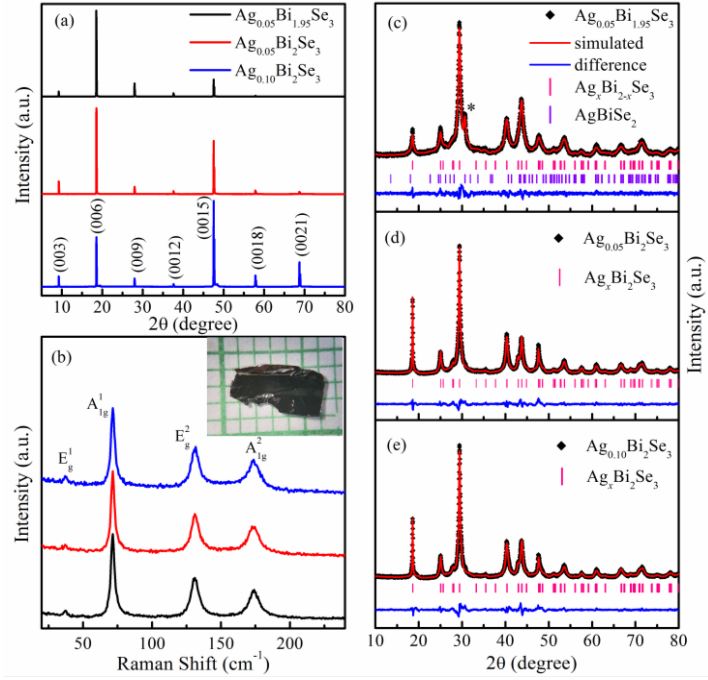

**Figure 1** Room temperature (a) X-ray diffraction pattern performed on flakes showing only (00*l*) peaks, (b) Raman spectra showing Raman active modes. Inset shows the image of cleaved flake. Rietveld refinements using *FullProf* Suite for (c) Ag$_{0.05}$Bi$_{1.95}$Se$_3$, (d) Ag$_{0.05}$Bi$_2$Se$_3$ and (e) Ag$_{0.10}$Bi$_2$Se$_3$.

**Table 1.** Lattice parameters for Ag$_x$Bi$_{2-x}$Se$_3$ ($x$ = 0.05), and Ag$_x$Bi$_2$Se$_3$ ($x$ = 0.05, 0.10) after Rietveld refinement fit.

| Samples | a (Å) | c (Å) | V (Å$^3$) | c/a |
|---|---|---|---|---|
| Ag$_{0.05}$Bi$_{1.95}$Se$_3$ | 4.1376(4) | 28.647(5) | 424.72(9) | 6.923 |
| Ag$_{0.05}$Bi$_2$Se$_3$ | 4.1374(3) | 28.633(2) | 424.47(5) | 6.921 |
| Ag$_{0.10}$Bi$_2$Se$_3$ | 4.1352(3) | 28.620(3) | 423.84(6) | 6.921 |

Figure 2 depicts the field-emission scanning electron microscopy (FESEM) image, showing the layered morphology of the samples and chemical composition calculated from EDS spectra are shown in Table 2. This implies that Ag is contained in the samples, although it is less than the nominal value taken during sample synthesis.

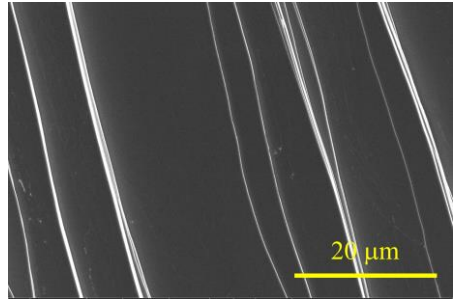

**Figure 2** FESEM image on one of the samples.

**Table 2.** Elemental composition of compounds calculated from EDS spectra averaged over 5 different spots.

| Sample | Ag (at. %) | Bi (at. %) | Se (at. %) | Calculated composition |
|---|---|---|---|---|
| $Ag_{0.05}Bi_{1.95}Se_3$ | 0.33 | 35.72 | 63.94 | $Ag_{0.02}Bi_{1.95}Se_{3.5}$ |
| $Ag_{0.05}Bi_2Se_3$ | 0.71 | 35.50 | 63.79 | $Ag_{0.04}Bi_2Se_{3.6}$ |
| $Ag_{0.10}Bi_2Se_3$ | 0.85 | 35.50 | 63.64 | $Ag_{0.05}Bi_2Se_{3.6}$ |

## 2.2 Electrical Transport

Figure 3 shows the temperature dependence of longitudinal resistivity ($\rho_{xx}$), normalised with value at 300 K at different magnetic fields (0, 2, 5 and 8 T). The temperature variation of resistivity for Ag-doped compounds (Fig. 3 (b, c, d)) follows the metallic behaviour in the temperature range 1.8 - 300 K. $Ag_{0.05}Bi_{1.95}Se_3$ compound (Fig. 3(b)) show a clear upturn in resistivity values below ~ 50 K, and it almost saturates below 20 K (Inset in Fig. 3(b)). Such kind of upturn at low temperatures has been observed earlier in low carrier density pristine $Bi_2Se_3$ crystals ($n_H$ ~ $10^{16}$-$10^{17}$ cm$^{-3}$).[22]

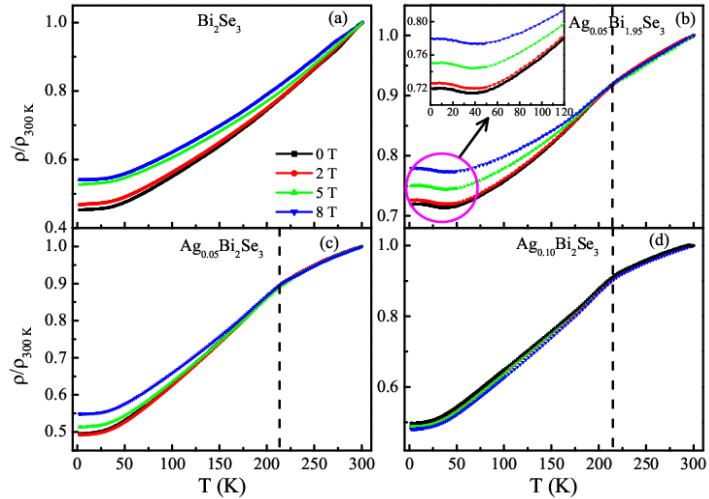

**Figure 3** Temperature dependence of longitudinal resistivity normalised at 300 K taken at different fields (0, 2, 5 and 8 T) for (a) $Bi_2Se_3$, (b) $Ag_{0.05}Bi_{1.95}Se_3$, (c) $Ag_{0.05}Bi_2Se_3$ and (d) $Ag_{0.10}Bi_2Se_3$. Inset in (b) panel shows the low temperature behaviour for the $Ag_{0.05}Bi_{1.95}Se_3$ compound.

It is to note that Hall resistivity for this compound is of the order of $10^{18}$ cm$^{-3}$ and does not change much at low-temperature range, suggesting that carrier density remains constant in this regime. Thus, we attribute this upturn due to possible effect of secondary phase $AgBiSe_2$ in $Ag_{0.05}Bi_{1.95}Se_3$ than $Ag_xBi_2Se_3$ in $Bi_2Se_3$.[23] Furthermore, the observed slope change in resistivity behaviour of Ag-doped compounds is quite different as compared to pristine $Bi_2Se_3$ in the temperature range 200 - 300 K. With the application of magnetic fields, resistivity increases for the $Bi_2Se_3$, $Ag_{0.05}Bi_2Se_3$ and $Ag_{0.05}Bi_{1.95}Se_3$ exhibiting positive MR but a negligible change in $Ag_{0.10}Bi_2Se_3$. The resistivity values in presence of magnetic fields does not enhance much above 200 K as compared to low temperature values, where a clear increase has been observed.

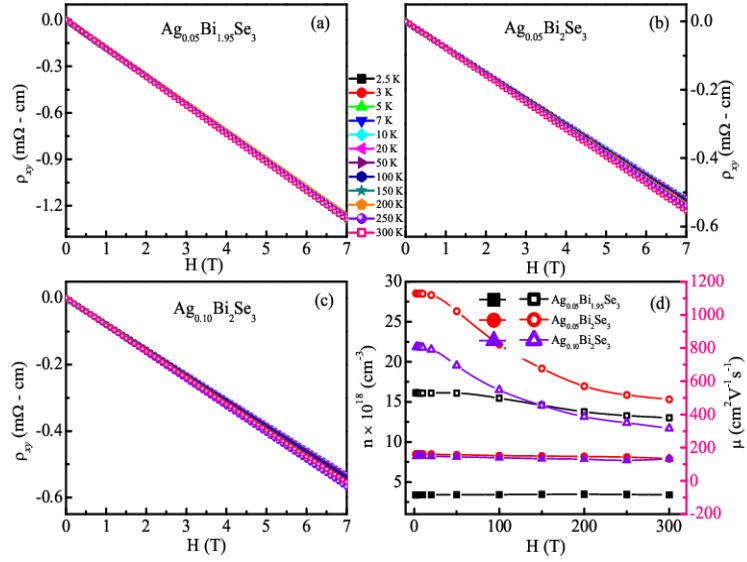

**Figure 4** (a), (b), (c) Magnetic field dependence of Hall resistivity ($\rho_{xy}$) at various temperatures for $Ag_{0.05}Bi_{1.95}Se_3$, $Ag_{0.05}Bi_2Se_3$, $Ag_{0.10}Bi_2Se_3$ respectively, and (d) shows variation of carrier density, $n_H$ (close symbol) and Hall mobility, $\mu_H$ (open symbol) as a function of temperature.

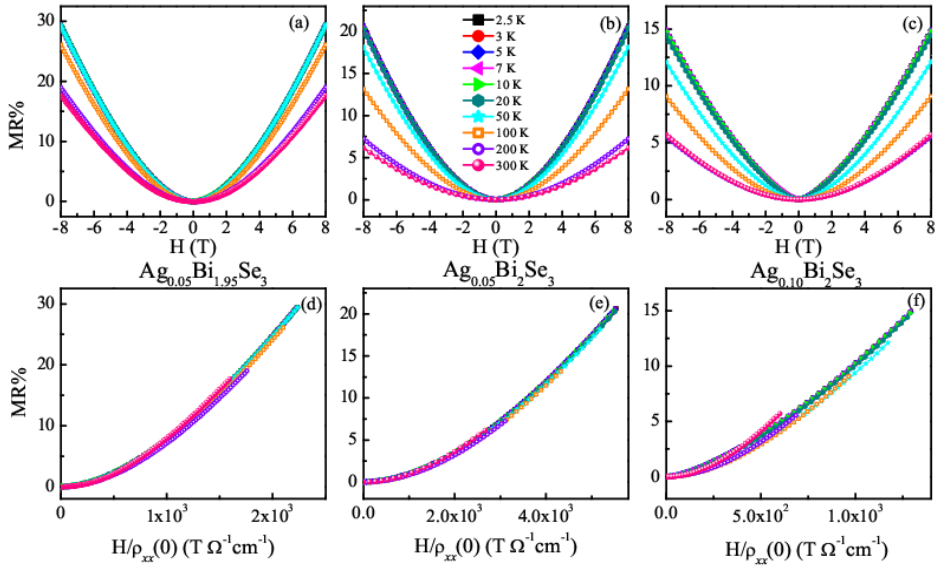

**Figure 5** (Top panel) Magnetoresistance (MR) as a function of magnetic field (H) at different temperatures. (Bottom panel) Kohler's plot for the samples: (a, d) $Ag_{0.05}Bi_{1.95}Se_3$, (b, e) $Ag_{0.05}Bi_2Se_3$, and (c, f) $Ag_{0.10}Bi_2Se_3$ at different temperatures.

Hall effect measurements revealed electrons as the majority charge carriers at all temperatures as shown in Fig. 4. Values of $n_H$ are determined from the slope ($R_H = \rho_{xy}/H$) of linear fits to magnetic field dependent Hall resistivity ($\rho_{xy}$) data using single band model. The carrier density is in the order of $10^{18}$ cm$^{-3}$ for all the compounds and found to be weakly temperature dependent in the range 1.8 - 300 K, which can also be seen from the Fig. 4 (a, b, c) that all the curves overlap onto each other at all temperatures. Hall mobility is estimated from the carrier density ($n_H$) and longitudinal resistivity ($\rho_{xx}$) at zero magnetic field, using the relation $1/\rho_{xx} = n_H e \mu_H$. The variation in mobility with the temperature is more pronounced than that in carrier density, owing to the temperature dependent resistivity. We find that $Ag_{0.05}Bi_{1.95}Se_3$ tend to decrease the mobility and carrier concentration (~ $3.4 \times 10^{18}$ cm$^{-3}$) as compared to $Ag_xBi_2Se_3$ which adds the carrier charges (~ $8 \times 10^{18}$ cm$^{-3}$), which are more than pristine $Bi_2Se_3$ (~ $2.4 \times 10^{18}$ cm$^{-3}$) at low temperature.[20]

Figure 5 (a, b, c) shows the change in isothermal resistivity in presence of magnetic fields plotted as magnetoresistance (MR = [$\rho_{xx}$(H) – $\rho_0$]/ $\rho_0$×100 %) at different temperatures. We observe a H$^2$ dependence at low fields and positive linear MR at high fields up to 8 T. MR does not show any sign of saturation up to magnetic fields of 8 T, which has been observed in wide range of topological materials such as Ag$_{2+\delta}$Se, Bi$_2$Se$_3$ and other 3D TIs.[24, 25] Generally, MR in metals saturate at high magnetic fields once the condition $\omega_c\tau \gg 1$ or $\mu H > 1$ is reached so that Landau levels can be formed. The linear band dispersion of Dirac fermions leads to the non-saturating linear MR. The origin of non-saturating linear MR can be explained on the basis of Abrikosov's linear quantum MR theory developed for the extreme quantum limit ($\hbar\omega_c > E_F$) such that all electrons occupy the lowest Landau level. [26] Figure 5 (d, e, f) presents plots for the Kohler's scaling at different temperatures. Kohler's rule is derived from the semi-classical transport theory of Boltzmann equation.[27]] The field dependence of MR can be rescaled as magnetic field by zero-field resistivity ($\rho_0$) given by equation $MR = \Delta\rho/\rho_0 = f(\omega\tau) = f(H/\rho_0)$.[28] The field dependence of resistivity is included in the quantity $\omega\tau$, where $\tau$ is the scattering rate ($\rho_0$ = m/ne$^2$) and $\omega$ is the cyclotron frequency at which magnetic field causes the charge carrier to sweep across the Fermi surface. The scaling of Kohler's rule holds if there is single type of charge carrier and scattering rate is similar at all points of Fermi surface. As the plots of $\Delta\rho/\rho_0$ as a function of $H/\rho_0$ at distinct temperatures fall onto a single curve.  Kohler's rule is followed in all Ag-doped samples studied here.

**ARPES**

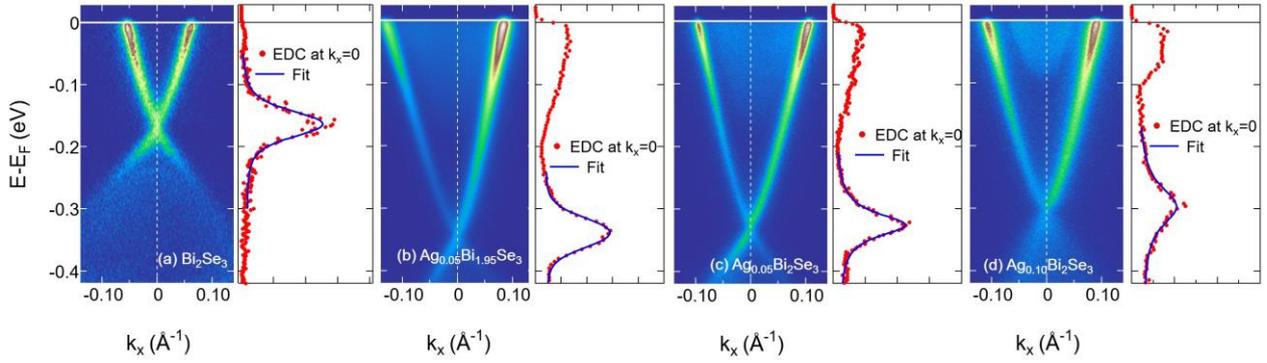

**Figure 6** ARPES spectra with the energy distribution curves (EDC) at the $\bar{\Gamma}$ point ($k_x$= 0) for Bi$_2$Se$_3$, Ag$_{0.05}$Bi$_{1.95}$Se$_3$, Ag$_{0.05}$Bi$_2$Se$_3$ and Ag$_{0.10}$Bi$_2$Se$_3$.

ARPES experiments were performed to probe the electronic band structure which show the characteristic surface states on the bulk Bi$_2$Se$_3$ and Ag doped Bi$_2$Se$_3$ single crystals. Figure 6 show the ARPES spectra for different crystals revealing the presence of topological surface states (TSS), bulk valence band (BVB) as a function of Fermi wave vector ($k_F$) measured along $\bar{\Gamma} - \bar{M}$ high symmetry direction. We observed that surface states band cross $E_F$ and intersects at the Dirac point ($E_D$) located at a binding energy ($E_B$) of ~ -180 meV for Bi$_2$Se$_3$, shifts to ~ -350 meV for both Ag$_{0.05}$Bi$_{1.95}$Se$_3$ and Ag$_{0.05}$Bi$_2$Se$_3$ and ~ -300 meV for Ag$_{0.10}$Bi$_2$Se$_3$. Thus, it shows the downward shift of Dirac point upon Ag doping in Bi$_2$Se$_3$. This kind of shift has been earlier observed in Nb$_x$Bi$_2$Se$_3$ ($x$ = 0.2, 0.25), Sr$_x$Bi$_2$Se$_3$ (0 ≤ $x$ ≤ 0.12), Cu$_x$Bi$_2$Se$_3$ (0 ≤ $x$ ≤ 0.25).[9, 10, 29, 30] It further reveals that Ag doping in Bi$_2$Se$_3$ does not alter the surface band dispersion. The Fermi level crosses the bulk conduction band which indicates that crystals are *n*-type, in agreement with our Hall resistivity data. The parameters presented in Table 3 are extracted from momentum distribution curves (MDC) at the $\bar{\Gamma}$ point fitted by a single Lorentzian function that implies single peak structure suggesting gapless states at the Dirac point. The Fermi wavevector ($k_F$) values are extracted from the analysis of the MDC of the ARPES spectra by fitting the MDC at the Fermi level using Voigt function. Using $k_F$, we have estimated the charge carrier density $n_{2D,TSS} = \frac{\pi k_F^2}{(2\pi)^2}$ for the TSSs. The Fermi velocity ($v_F$) and effective mass ($m^*$) were calculated using the formula $v_F = \frac{1}{\hbar}\left(\frac{\partial E}{\partial k}\right)_{k=k_F}$ and $m^* = \frac{\hbar k_F}{v_F}$ respectively. The linewidth $\Gamma = (\Delta E) = \left(\frac{\partial E}{\partial k}\right) \times (\delta k)$ and relaxation time $\tau = \frac{\hbar}{\Gamma}$ are extracted using the MDC width $\delta k$.[31-33] Lorentzian line width ($\delta k$) from the Voigt function  (we assumed the Gaussian line width represents the instrumental resolution) is used for the evaluation of $\Gamma$ and $\tau$.

**Table 3** Parameters obtained from ARPES measurements for $Bi_2Se_3$, $Ag_{0.05}Bi_{1.95}Se_3$, $Ag_{0.05}Bi_2Se_3$, $Ag_{0.10}Bi_2Se_3$.

| Parameters | $Bi_2Se_3$ | $Ag_xBi_{2-x}Se_3$ | $Ag_xBi_2Se_3$ | |
|---|---|---|---|---|
| | | $x = 0.05$ | $x = 0.05$ | $x = 0.10$ |
| $k_F$ ($10^6$ cm$^{-1}$) | 6.24 | 10.74 | 10.05 | 9.95 |
| $A_F$ ($10^{13}$ cm$^{-2}$) | 12.22 | 36.27 | 31.7 | 31.08 |
| $n_{2D,TSS}$ ($10^{12}$ cm$^{-2}$) | 3.10 | 9.19 | 8.04 | 7.88 |
| $v_F$ ($10^7$ cm s$^{-1}$) | 4.28 | 4.85 | 4.94 | 4.65 |
| $m^*$ ($m_e$) | 0.16 | 0.25 | 0.26 | 0.24 |
| $\Gamma$ (meV) | 11 | 24 | 19 | 39 |
| $\tau$ ($10^{-14}$ sec) | 5.9 | 2.7 | 3.5 | 1.6 |
| MDC linewidth: $\delta k$ (Å$^{-1}$) | 0.004 | 0.007 | 0.006 | 0.0128 |
| Mean free path $l$ (nm) | 25 | 13 | 17 | 8 |

Figure 7 represents the set of constant energy contours of the Dirac cone at different energies for $Bi_2Se_3$ and $Ag_{0.10}Bi_2Se_3$ samples. The evolution of circular Fermi surface height of $E_F$ is referenced to the Dirac point. The shape of the constant energy contour is nearly circular near the Dirac point, with increasing Fermi surface area above 180 meV. However, effect of hexagonal deforming has been reported earlier in $Bi_2Se_3$, which is negligible in our sample.[34]

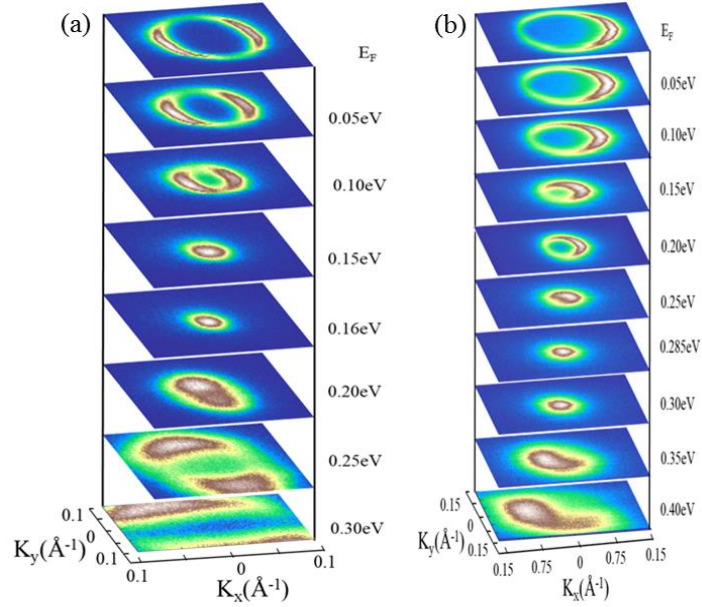

**Figure 7** A set of constant energy contours at different binding energies for (a) $Bi_2Se_3$ and (b) $Ag_{0.10}Bi_2Se_3$.

## 3. Conclusion

We have investigated magnetotransport of Ag-doped $Bi_2Se_3$ crystals synthesized using different methods, such that in one case Ag-substitution at Bi sites is more probable. The difference has been observed as an upturn in low-temperature resistivity behavior for $Ag_{0.05}Bi_{1.95}Se_3$ which is different from $Ag_xBi_2Se_3$. All compounds exhibit quadratic field dependence of the MR and follows Kohler's rule, implying the dominance of bulk transport with single type of scattering mechanism in these systems. The carrier concentration in Ag doped compounds were found to 10 – 80 times larger than for $Bi_2Se_3$. The downward shift in the Dirac point energy has been observed in Ag-doped sample using ARPES. The observed shift is consistent with the doping of the electronic charge carriers. The Dirac point is found to shift by ~ 120 meV upon Ag doping. The ARPES results do not show any specific difference in the compounds prepared with two different processes.

## 4. Experimental details

Single crystals of the compounds $Ag_xBi_2Se_3$ ($x$ = 0.05, 0.10) and $Ag_xBi_{2-x}Se_3$ ($x$ = 0.05) have been synthesised using the melt-grown technique. Stoichiometric amounts of Bi granules (≥ 99.99 %), Se granules < 5 mm (≥ 99.999 %), Ag granules ~ 250 μm (≥ 99.99 %) from *Sigma-Aldrich* were taken in sealed evacuated quartz tubes. For $Ag_xBi_2Se_3$ ($x$ = 0.05, 0.10), firstly $Bi_2Se_3$ was prepared for all the compositions. The compounds were heated at 850 °C for 24 hrs, followed by cooling to 550 °C at the rate 10 °C /h, where they were kept for 72 hrs. Then, the compounds were furnace off cooled to room temperature. Second, the grown $Bi_2Se_3$ and Ag were kept at 850 °C for 48 hrs, followed by slow cooling (3 °C /h) to 400 °C, then furnace off cooled to room temperature. For $Ag_xBi_{2-x}Se_3$, elemental Bi, Ag, and Se were accurately weighed according to the stoichiometric ratios and vacuum-sealed in quartz tubes. These were kept at 850 °C for 48 hrs, cooled (10 °C /h) to 550 °C, where they were kept for 72 hrs. Finally, for better homogeneity of materials, all the compounds were again mixed thoroughly using mortar and pestle, pelletized and vacuum sealed and kept at 850 °C for 48 hrs, then slow cooled (3 °C /hrs) to 400 °C, and furnace off cooled to room temperature. The bulk crystals were then cleaved easily using scalpel blade to obtain thin, flat, and shiny flakes.

Phase confirmations were studied using powder X-ray diffraction using a Rigaku SmartLab rotating anode diffractometer with Cu-Kα (λ= 1.5418 Å) at room temperature. Raman spectra was obtained using Horiba LabRAM HR Evolution Raman spectrometer equipped with a 532 nm laser. The elemental composition and surface morphology of samples have been determined using the energy-dispersive X-ray spectroscopy (EDS) equipped with the field-emission scanning electron microscopy (FESEM, Nova Nano SEM-450, FEI). EDS measurements were performed at different spots to estimate the elemental composition in the samples. Magnetotransport measurements were carried out in a DynaCool PPMS (*Quantum Design*). ARPES measurements were performed using μ-Laser ARPES system[35] with a base pressure of $5 \times 10^{-9}$ Pa at $T$ = 20 K. The measurement details are same as discussed elsewhere.[36]

**Acknowledgement**

SS acknowledges IIT Mandi for HTRA fellowship. CSY acknowledges the experimental facility at AMRC, IIT Mandi. The ARPES measurements were performed with the approval of the Proposal Assessing Committee of the Hiroshima Synchrotron Radiation Center (Proposal Number: 19BU009). We thank N-BARD, Hiroshima University for supplying the liquid Helium.

**Conflict of interest**

The authors declare no conflict of interest.